\documentclass[a4paper]{jpconf}
\usepackage{graphicx}
\begin{document}
\title{Pion production in neutrino-nucleus collisions}

\author{E. Hern\'andez}

\address{Departamento
de F\'\i sica Fundamental e IUFFyM, Universidad de Salamanca, E-37008
Salamanca, Spain}

\ead{gajatee@usal.es}
\author{J. Nieves}
\address{Instituto de F\'\i sica Corpuscular (IFIC), Centro Mixto
  CSIC-Universidad de Valencia, Institutos de Investigaci\'on de
  Paterna, Apartado 22085, E-46071 Valencia, Spain}

\ead{jmnieves@ific.uv.es}
\author{M.J. Vicente Vacas}
\address{Departamento de F\'\i sica Te\'orica e IFIC, Centro Mixto
  Universidad de Valencia-CSIC, Institutos de Investigaci\'on de
  Paterna, Apartado 22085, E-46071 Valencia, Spain}
  \ead{Manuel.J.Vicente@uv.es}

\begin{abstract}
We compare our pion production results with recent MiniBooNE data measured in
mineral oil. 
Our total cross sections lie below  experimental data  for  neutrino
energies above 1\,GeV.  Differential cross sections show our model produces too
few high energy pions in the forward direction as compared to data. The 
agreement with experiment 
improves by artificially removing 
pion final state interaction.
\end{abstract}

\section{Introduction}
In this contribution we  present our results~\cite{Hernandez:2013jka} for 
$\nu_\mu/\bar\nu_\mu$-induced one pion production
cross sections in mineral oil ($CH_2$) for neutrino
 energies
below 2\,GeV. These results are compared to the experimental data 
obtained by the MiniBooNE 
Collaboration~\cite{AguilarArevalo:2010bm,AguilarArevalo:2010xt,
AguilarArevalo:2009ww}.

Our calculational starting point is the pion
production model at the nucleon level of
Refs.~\cite{Hernandez:2007qq,Hernandez:2010bx}, that we have
 extended from the $\Delta(1232)$ region  up to
2\,GeV neutrino energies by the inclusion of the $D_{13}(1520)$ 
 resonance. Apart from the $\Delta(1232)$ already present in the model, 
 the $D_{13}(1520)$ resonance gives the most important
 contribution in that extended energy region~\cite{Leitner:2008ue}.  In-medium 
 corrections in the production process  include 
 Pauli-blocking, Fermi motion, and  the modification of the $\Delta$ resonance 
 properties 
inside the nuclear medium. Not only  the $\Delta$
propagator is modified, but there is also a new pion production contribution
(referred to as $C_Q$ in the following) that
comes from the changes in the $\Delta$ width   in
the nuclear environment.  For  pion final state interaction (FSI) we use a
cascade program that follows  Ref.~\cite{Salcedo:1987md} where a general simulation code for inclusive
  pion nucleus
reactions was developed. When  coherent pion
production  is 
possible we evaluate its contribution  using the model in 
Refs.~\cite{Amaro:2008hd,Hernandez:2010jf}.
% but with the nucleon-to-Delta form factors as 
%extracted in
%  Ref.~\cite{Hernandez:2010bx}. 
Due to lack of space, here we shall  just show the
  results. For details we refer the reader to Ref~\cite{Hernandez:2013jka}.
Our results  are qualitatively similar to those obtained by other 
groups~\cite{Lalakulich:2012cj,Sobczyk.:2012zj}. 
\section{Results and
comparison with MiniBooNE data}
\label{sec:results}

\begin{figure}[tbh]
 \begin{center}
 \includegraphics[height=5.cm]{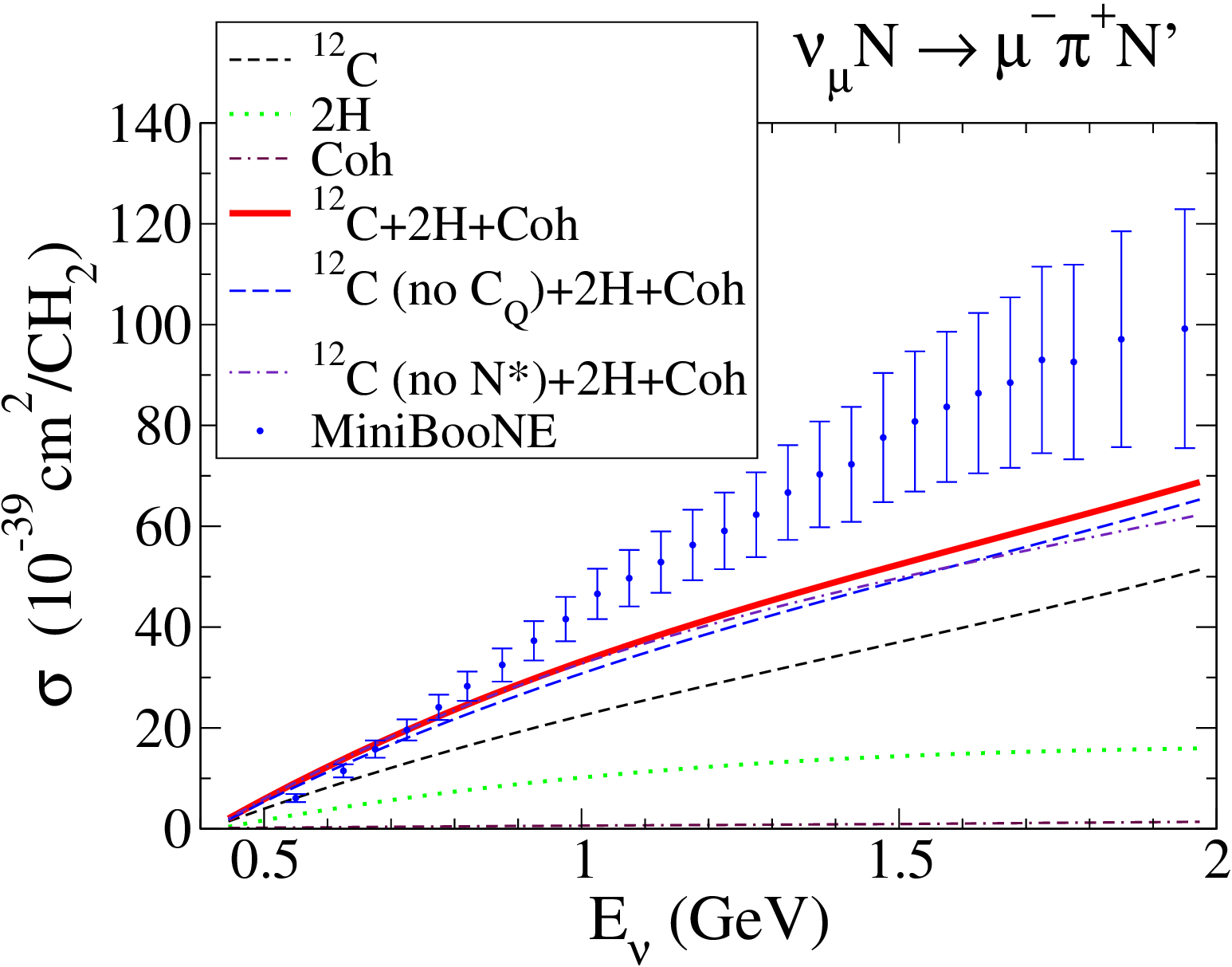}\hspace{1.5cm}
\includegraphics[height=4.85cm]{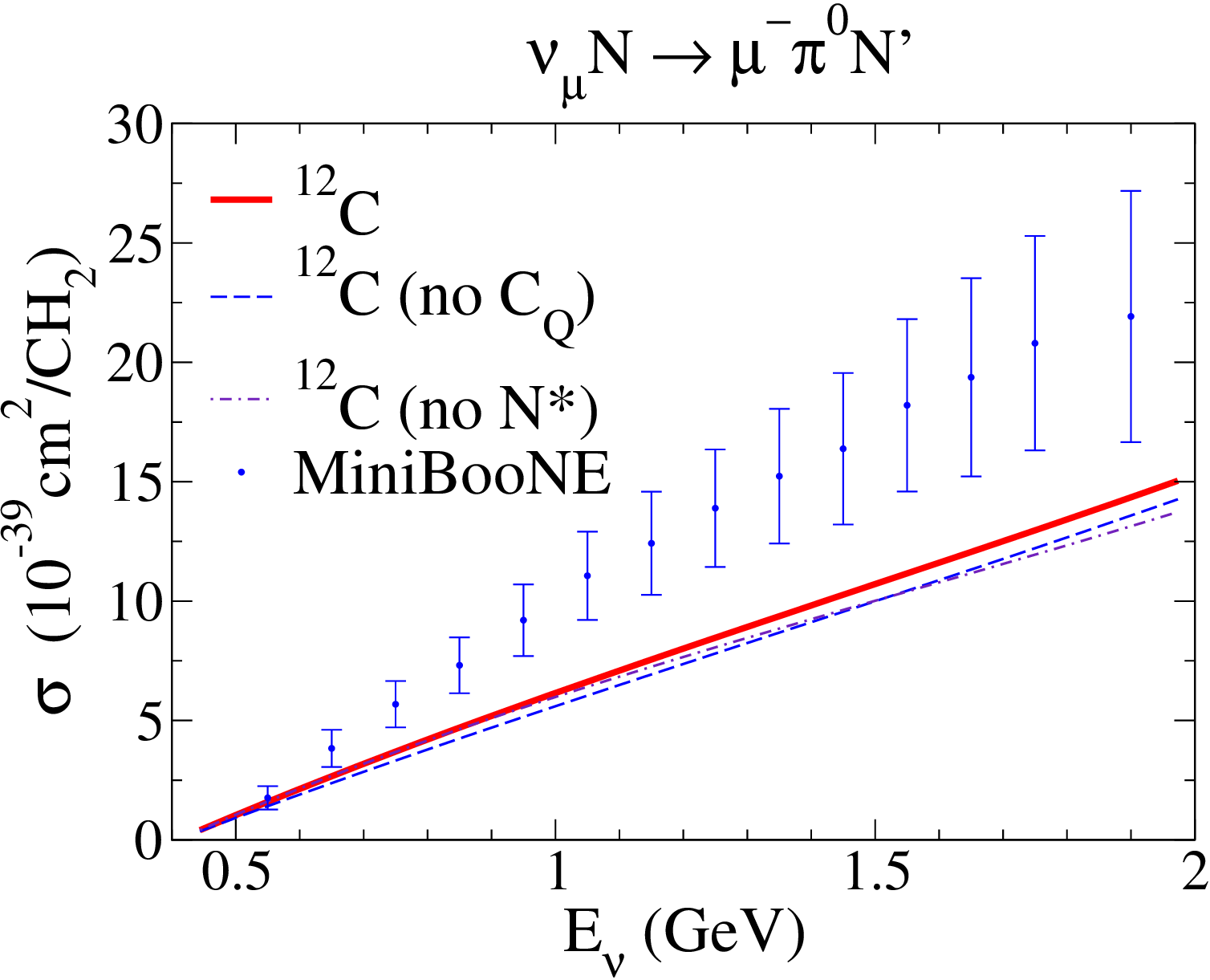}
\caption{ $1\pi^+$ (left) and $1\pi^0$ (right) total production cross section 
for $\nu_\mu$ $CC$ interaction
in mineral oil.  Dashed line:
$^{12}C$ contribution. Dotted line: $H_2$
contribution. Double-dashed dotted line: Coherent contribution.
Solid line: Total contribution. Broken line: Same as solid line but without the
$C_Q$ contribution. Dashed-dotted line: Same as solid line but without the
contribution from the $D_{13}$.  
 Experimental data taken from
Refs.~\cite{AguilarArevalo:2010bm}  and 
\cite{AguilarArevalo:2010xt}.}
  \label{fig:totalccpip}
  \end{center}
\end{figure}

 We start by showing total cross sections for a given neutrino energy. In the 
 left panel of Fig.~\ref{fig:totalccpip} we compare with 
  MiniBooNE data our results for 
 $\pi^+$ production in a charged current ($CC$) process.  
Our cross sections are below data for neutrino energies above $0.9$\,GeV.
The contribution from the $D_{13}$ resonance only plays a role above 
$E_\nu=1.2\,$GeV, making some 8\% of the
total at the highest neutrino energy. The $C_Q$ term contributes 
for all energies, 
being   around 8\% of the total. Similar
results are obtained for  a final $\pi^0$ (right panel). 
 \begin{figure}[tbh]
 \begin{center}
 \includegraphics[height=4.5cm]{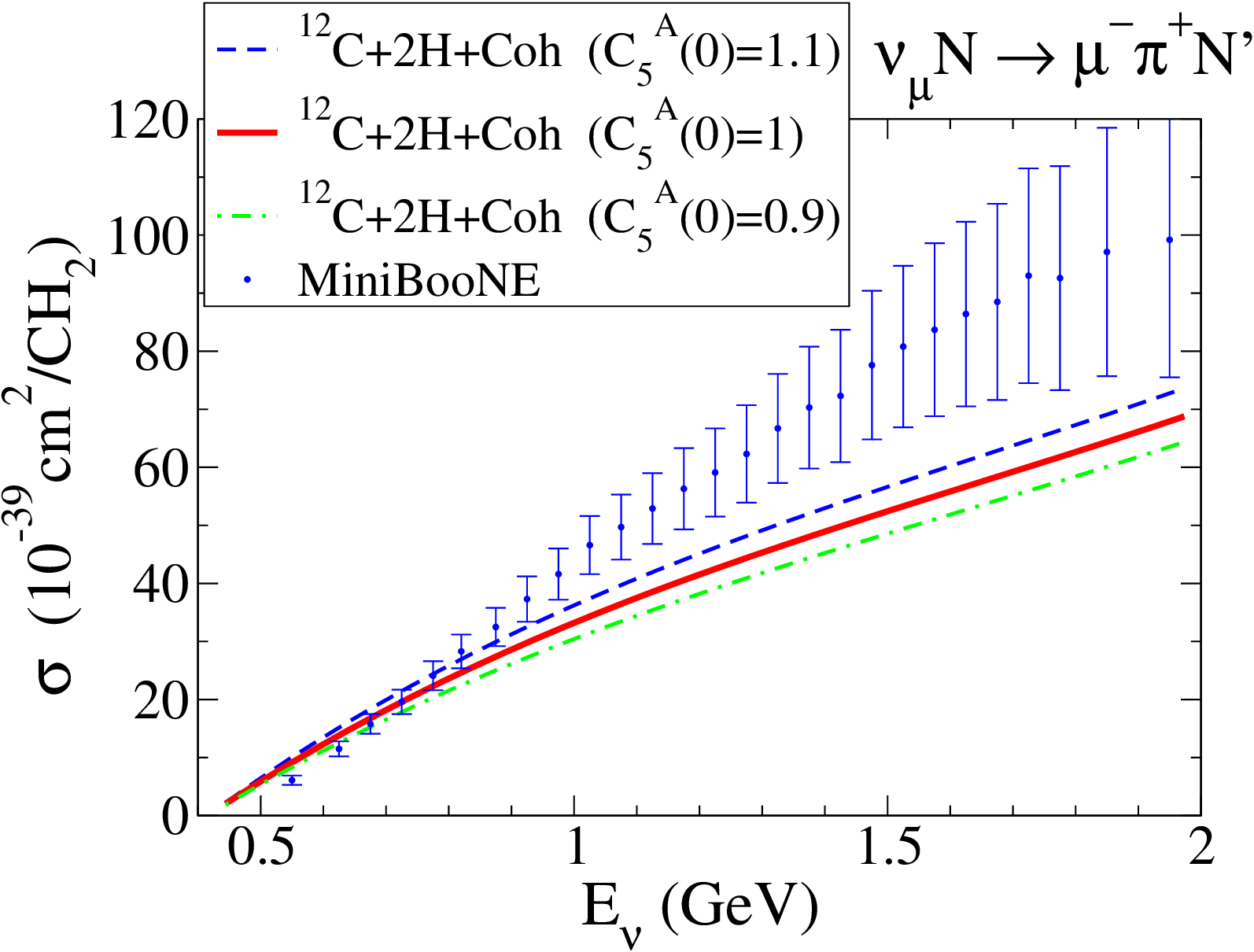}
\caption{ $1\pi^+$ total production cross section for
 CC interaction in mineral oil. Solid line: Our full model with
$C^A_5(0)=1$. Dashed line: Full model with $C^A_5(0)=1.1$. Dashed-dotted
line: Full model with $C^A_5(0)=0.9$.
 Experimental data taken from
Ref.~\cite{AguilarArevalo:2010bm}.}
  \label{fig:totalccpipca5}
  \end{center}
\end{figure}

In Fig.~\ref{fig:totalccpipca5} we show the effects in our results of changing
the value of the dominant axial nucleon-to-Delta form factor within  
the uncertainties in its determination in Ref.~\cite{Hernandez:2010bx}. A larger value than the central one we use
($C^A_5(0)=1$) seems to be preferable in the high energy region.
\begin{figure}[tbh]
\begin{center}
 \includegraphics[height=4.5cm]{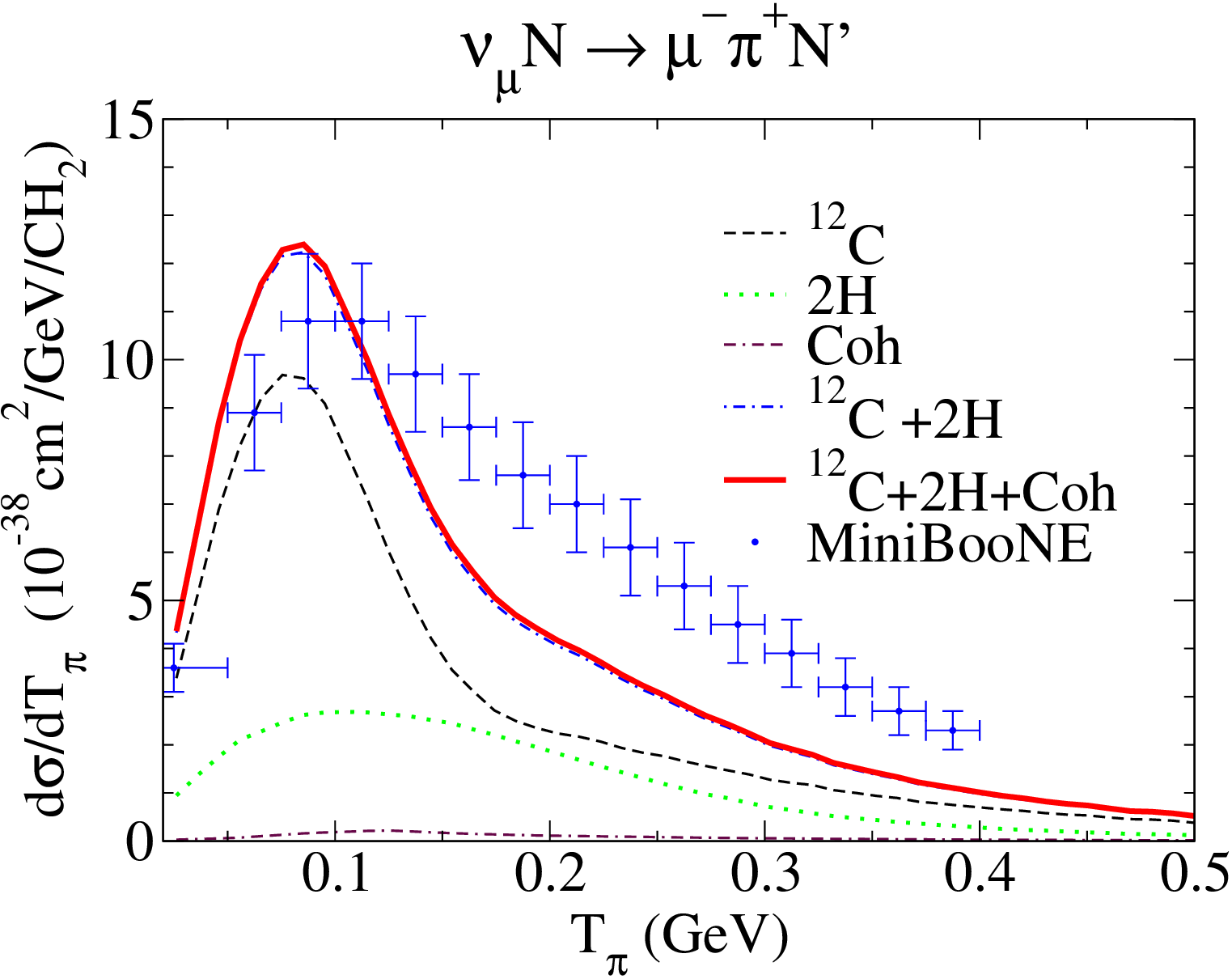}\hspace{1.5cm}
\includegraphics[height=4.5cm]{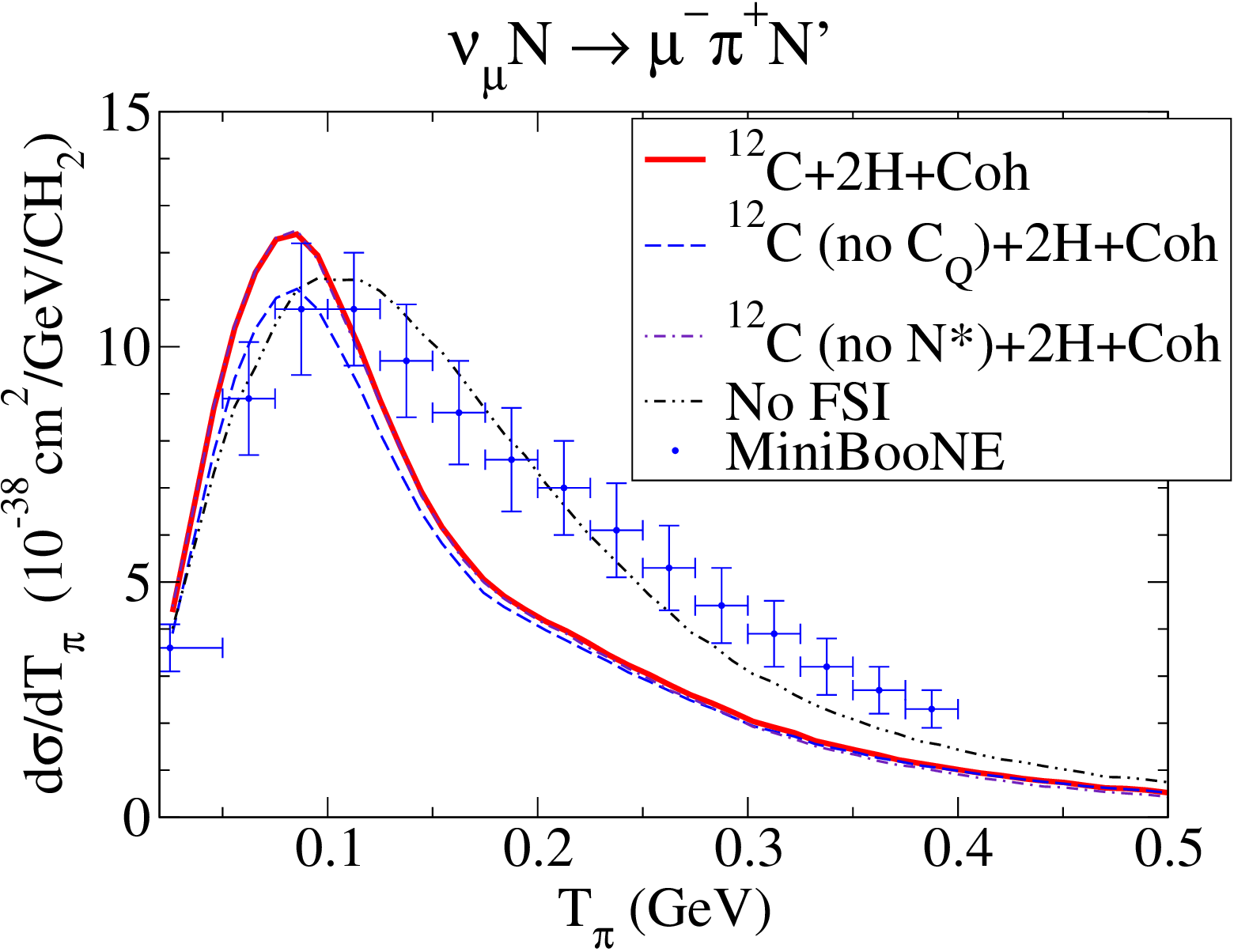}
\caption{ Differential $\frac{d\sigma}{dT_\pi}$ 
cross section for charged current $1\pi^+$ production by $\nu_\mu$ in mineral
oil. Captions as in Fig.~\ref{fig:totalccpip}. We also show in the right panel
results without FSI of the final pion (double-dotted dashed line).  Data
 from Ref.~\cite{AguilarArevalo:2010bm}. }
  \label{fig:cctpi}
  \end{center}
\end{figure}
\begin{figure}[tbh]
\begin{center}
 \includegraphics[height=4.5cm]{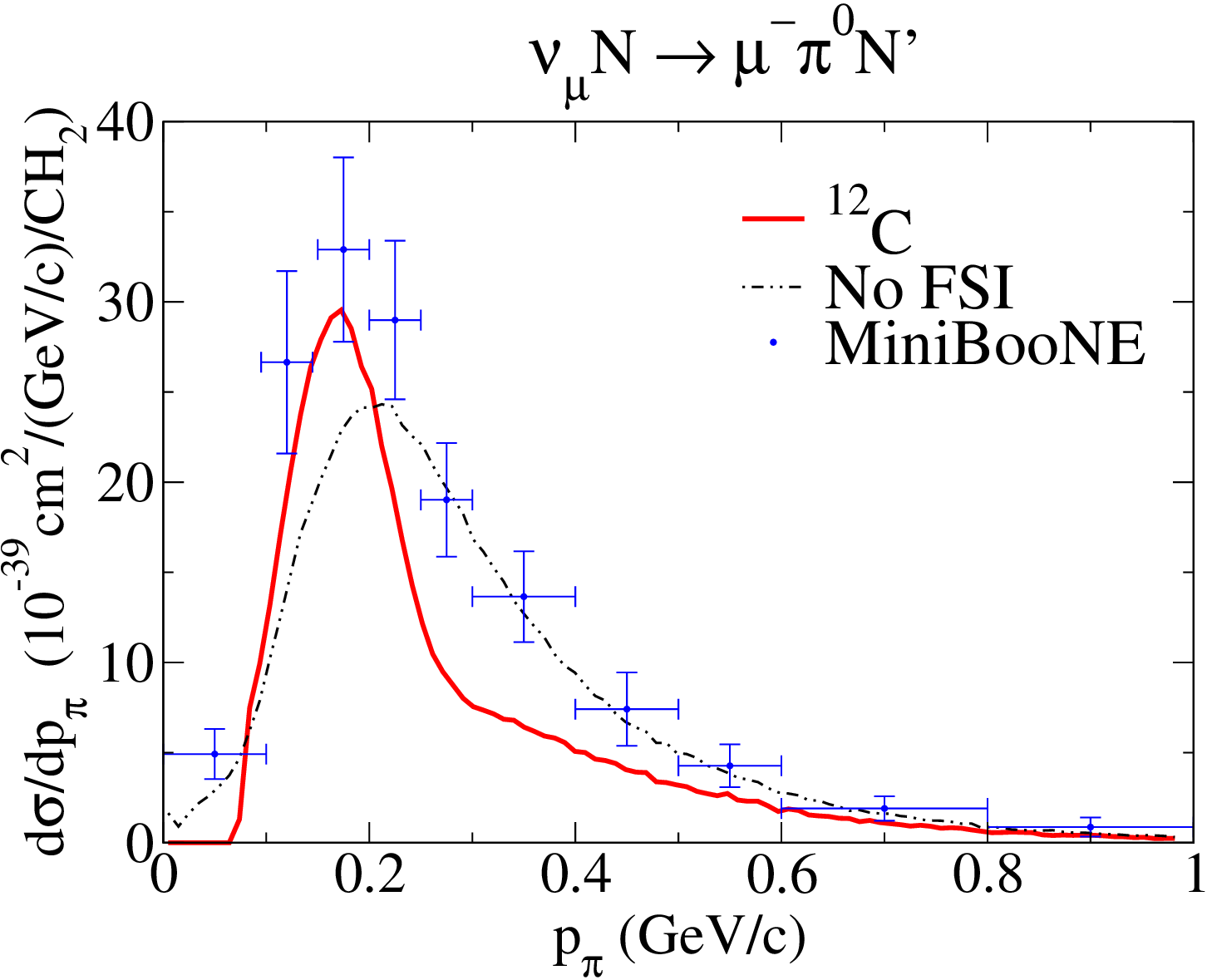}\hspace{1.5cm}
\includegraphics[height=4.5cm]{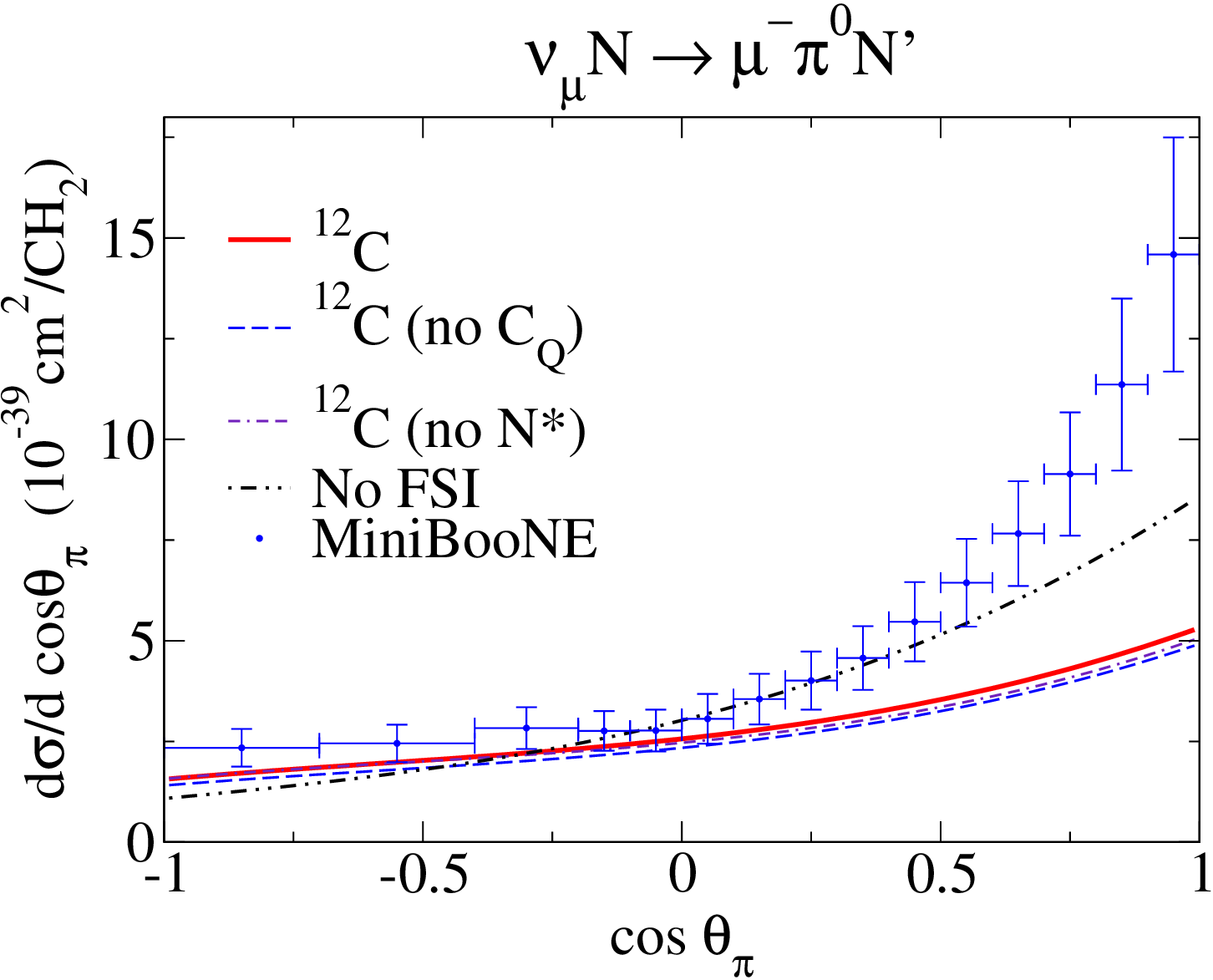}
\caption{ Differential $\frac{d\sigma}{dp_\pi}$ (left panel) and
$\frac{d\sigma}{d\cos\theta_\pi}$ (right panel)
cross section for $CC$ $1\pi^0$ production by $\nu_\mu$ in mineral
oil. Captions as in Figs.~\ref{fig:totalccpip} and \ref{fig:cctpi}. Data
 from Ref.~\cite{AguilarArevalo:2010xt}. }
  \label{fig:ccpi0}
  \end{center}
\end{figure}

In Fig.~\ref{fig:cctpi} we  compare results,  convoluted with the neutrino
 flux  in  
 Ref.~\cite{AguilarArevalo:2010bm}, for 
the differential $\frac{d\sigma}{dT_\pi}$ 
cross section for $CC$ $1\pi^+$ production by $\nu_\mu$. 
 We disagree with data for
 $T_\pi$ above 0.15\,GeV. The agreement improves if we 
 artificially remove FSI (see right panel).
  Also in the right panel we  show the effects of not including
 the $C_Q$ or $D_{13}$ contributions. By neglecting  the $C_Q$
 contribution the cross section decreases by some 10\% around the
 peak at $T_\pi=0.08\,$GeV. The $D_{13}$ plays  a very minor
  role since   the neutrino flux  peaks at around 600\,MeV.

Differential $\frac{d\sigma}{dp_\pi}$ 
and  $\frac{d\sigma}{d\cos\theta_\pi}$ 
cross sections for $CC$ $1\pi^0$ production by $\nu_\mu$ are shown in
Fig.\ref{fig:ccpi0}. For their evaluation we take the neutrino flux from
 Ref.~\cite{AguilarArevalo:2010xt}.  Our model agrees with data for pion momentum below 0.2\,GeV/c
but it produces too
few pions in the momentum region from 0.22 to 0.55\,GeV/c. As seen from the
angular distribution  those missing pions mainly go in the forward 
direction. The effects of ignoring FSI are also shown in both panels.
\begin{figure}[tbh]
\begin{center}
 \includegraphics[height=4.6cm]{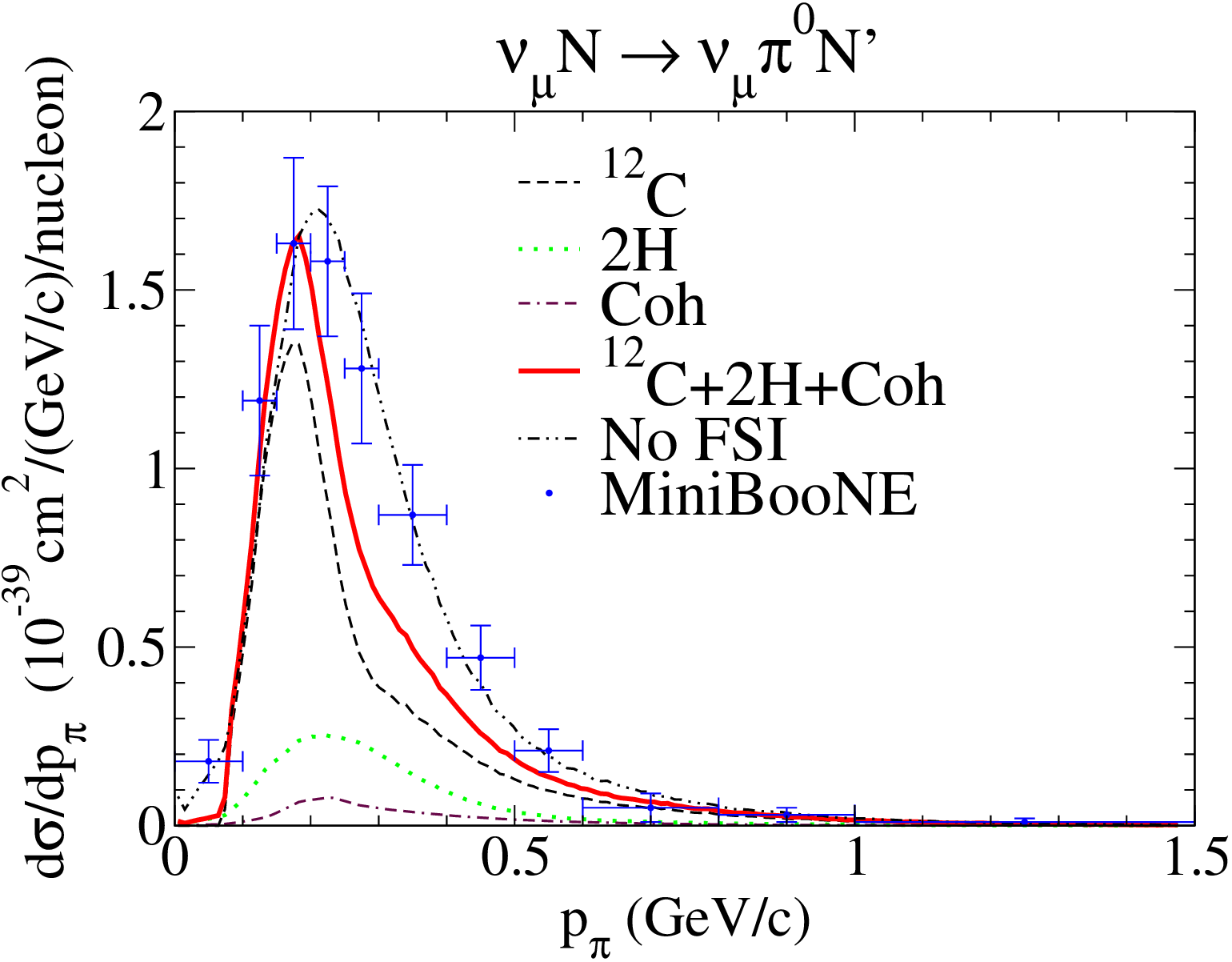}\hspace{1.5cm}
\includegraphics[height=4.6cm]{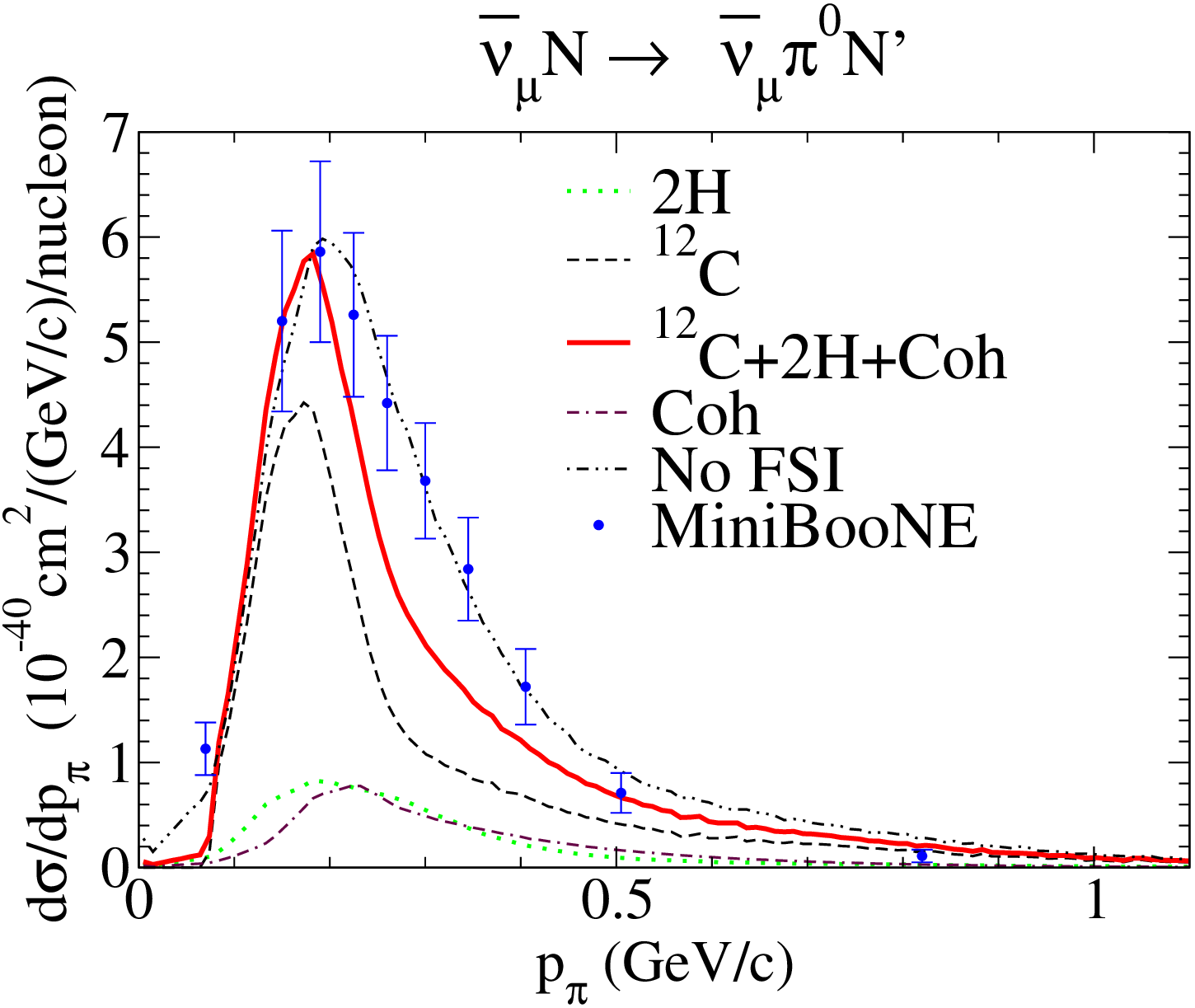}
\caption{ Differential $\frac{d\sigma}{dp_\pi}$ 
cross section per nucleon for $NC$ $1\pi^0$ production by $\nu_\mu$ (left panel)
and $\bar\nu_\mu$ (right panel) in mineral
oil. Captions as in Figs.~\ref{fig:totalccpip} and \ref{fig:cctpi}. Data
 from Ref.~\cite{AguilarArevalo:2009ww}. }
  \label{fig:ncmom}
  \end{center}
\end{figure}
\begin{figure}[tbh]
\begin{center} 
\includegraphics[height=4.6cm]{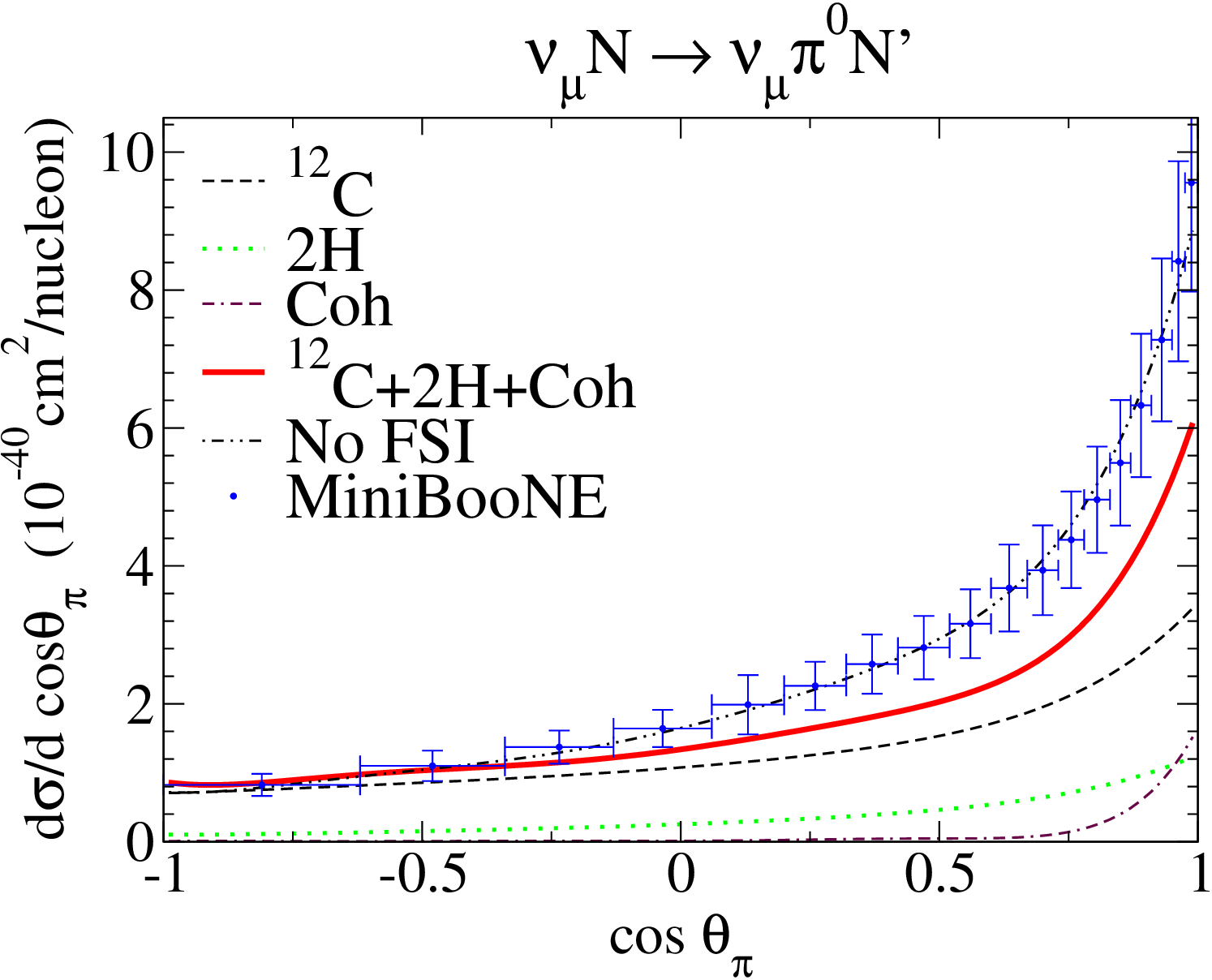}\hspace{1.5cm}
\includegraphics[height=4.6cm]{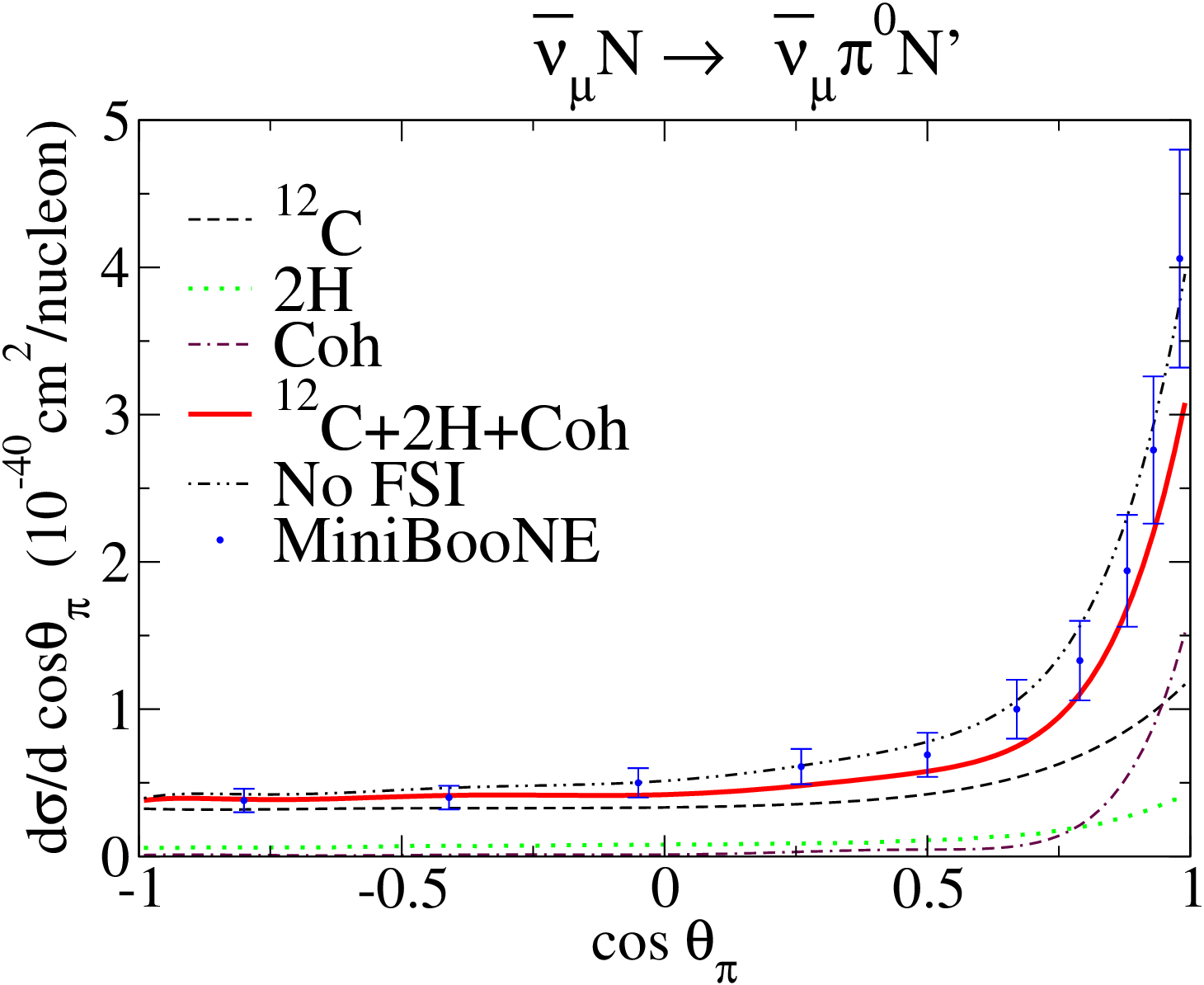}
\caption{ Differential $\frac{d\sigma}{d\cos\theta_\pi}$ 
cross section per nucleon for $NC$ $1\pi^0$ production.
 Captions as in Figs.~\ref{fig:totalccpip} and \ref{fig:cctpi}.  Data
 from Ref.~\cite{AguilarArevalo:2009ww}.}
  \label{fig:nccos}
  \end{center}
\end{figure}

In Figs.~\ref{fig:ncmom} and \ref{fig:nccos} we present the results for neutral current
 ($NC$) production  that we compare with data by the MiniBooNE 
 collaboration~\cite{AguilarArevalo:2009ww}. In each case we use the 
$\nu_\mu/\bar\nu_\mu$ fluxes reported in Ref.~\cite{AguilarArevalo:2009ww}. 
Fig.~\ref{fig:ncmom}  shows the
different contributions to the $\frac{d\sigma}{dp_\pi}$ differential cross
 section. Our 
 results show a depletion in the $0.25\sim0.5\,$GeV/c momentum 
region though the agreement is better than in the $CC$ case. 
The results agree with data if one neglects FSI.
Looking now at the
 differential $\frac{d\sigma}{d\cos\theta_\pi}$
cross sections shown in Fig.~\ref{fig:nccos}, one can see that 
our results
agree better with data in the antineutrino case where we are within
error bars except in the very forward direction.  A clear deficit in the
forward direction is seen for the reaction with neutrinos but the agreement 
is better than in 
the corresponding $CC$ reaction.  In both cases, the coherent contribution is
shown to be 
very relevant  in the forward direction. Once more, if one artificially switches off
FSI effects we get a good agreement  with data.

\section*{Acknowledgments}
 This research was supported by  the Spanish Ministerio de Econom\'{\i}a y 
 Competitividad and European FEDER funds
under Contracts Nos. FPA2010-
21750-C02-02, FIS2011-28853-C02-01, FIS2011-28853-C02-02  and  CSD2007-00042, 
by Generalitat
Valenciana under Contract No. PROMETEO/20090090
and by the EU HadronPhysics3 project, Grant Agreement
No. 283286.

\end{document}